\documentclass[a4paper]{article}

\usepackage{booktabs}
\usepackage{multirow}
\usepackage{graphicx}
\graphicspath{{figures/}}

\usepackage{stackrel}
\usepackage{xspace}
\newcommand{\ie}{i.\@\,e.,\@\xspace}
\newcommand{\eg}{e.\@\,g.,\@\xspace}

\newcommand{\etal}{et~al.\@\xspace}

\usepackage{INTERSPEECH2020}
\usepackage{footnote}
\usepackage{tablefootnote}

\usepackage{hyperref}

\usepackage{enumitem}

\usepackage[dvipsnames]{xcolor}

\let\footnote\thanks

\title{Detecting Adversarial Examples for Speech Recognition \\via Uncertainty Quantification}
\name{Sina D\"aubener$^1$\footnote{$^{1,2}$equal contribution}, Lea Sch\"onherr$^1$, Asja Fischer$^2$, Dorothea Kolossa$^2$
}

\address{
  Ruhr University Bochum, Germany\footnote{Funded by the Deutsche Forschungsgemeinschaft (DFG, German Research Foundation) under Germany's Excellence Strategy - EXC 2092 CASA - 390781972.}}
\email{\{sina.daeubener, lea.schoenherr, asja.fischer, dorothea.kolossa\}@rub.de}
\begin{document}

\maketitle
\begin{abstract}
Machine learning systems and also, specifically,  automatic speech recognition (ASR) systems are vulnerable against adversarial attacks, where an attacker maliciously changes the input. In the case of ASR systems, the most interesting cases are \emph{targeted} attacks, in which an attacker aims to force the system into recognizing given target transcriptions in an arbitrary audio sample. The increasing number of sophisticated, quasi imperceptible attacks raises the question of countermeasures. 

In this paper, we focus on hybrid ASR systems and compare four acoustic models regarding their ability to indicate uncertainty under attack: a feed-forward neural network and three neural networks specifically designed for uncertainty quantification, namely a Bayesian neural network, Monte Carlo dropout, and a deep ensemble.

We employ uncertainty measures of the acoustic model to construct a simple one-class classification model for assessing whether inputs are benign or adversarial. Based on this approach, we are able to detect adversarial examples with an area under the receiving operator curve score of more than 0.99. The neural networks for uncertainty quantification simultaneously diminish the vulnerability to the attack, which is reflected in a lower recognition accuracy of the malicious target text in comparison to a standard hybrid ASR system.

\end{abstract}

\section{Introduction}
\label{sec:intro}

An increasing number of smart devices are entering our homes to support us in our everyday life. Many of such devices are equipped with automatic speech recognition~(ASR) to make their handling even more convenient. While we rely on ASR systems to understand the spoken commands, it has been shown that adversarial attacks can fool ASR systems~\cite{carlini-18-audio, Schoenherr2019, alzantot-2018-did, schnoeherr-19-imperio}. These attacks add (to some extent) imperceptible noise to the original audio, which fools the ASR system to output a false---attacker-chosen---transcription.

This manipulated transcription can be especially dangerous in security- and safety-critical environments such as smart homes or self-driving cars. In such environments, audio adversarial examples may, for example, be used to deactivate alarm systems or to place unwanted online orders.

There have been numerous attempts to tackle the problem of adversarial examples in neural networks (NNs). However, it has been shown that the existence of these examples is a consequence of the high dimensionality of NN architectures~\cite{shamir,ilyas-19-bugs}. To defend against adversarial attacks, several approaches aim \eg at making their calculation harder by adding stochasticity and reporting prediction uncertainties~\cite{gal_adversarial, MNF_Louizos, feinman2017detecting}. Ideally, the model should display high uncertainties if and only if abnormal observations like adversarial examples or out-of-distribution data are fed to the system. 
Akinwande~\etal{}~\cite{akinwande-2020-anomaly-detection} and Samizade~\etal{}~\cite{samizade-2019-classification} used anomaly detection, either in the network's activations or directly on raw audio, to detect adversarial examples. However, both methods are trained for defined attacks and are therefore easy to circumvent~\cite{carlini_detection}.
Zeng~\etal{}~\cite{zeng-2018-multiversion} have combined the output of multiple ASR systems and calculated a \emph{similarity score} between the transcriptions. Nevertheless, due to the transferability-property of adversarial examples to other models, this countermeasure is not guaranteed to be successful~\cite{papernot-2016-transferability}.
Yang~\etal{}~\cite{yang-2018-characterizing} also utilize temporal dependencies of the input signal. For this, they compare the transcription of the entire utterance with a segment-wise transcription of the utterance. In case of a benign example, both transcriptions should be the same, which will typically not be the case for an adversarial example.
Other works leveraged uncertainty measures to improve the robustness of ASR systems in the absence of adversarial examples.
Vyas~\etal~\cite{vyas-2019-dropout} used dropout and the respective transcriptions to measure the reliability of the ASR system's prediction. Abdelaziz~\etal~\cite{abdelaziz-15-uncertainty} and Huemmer~\etal~\cite{huemmer-15-uncertainty} have previously utilized the propagation of observation uncertainties through the layers of a neural network acoustic model via Monte Carlo sampling to increase the reliability of these systems under acoustic noise.

We combine the insights about uncertainty quantification from the deep learning community with ASR systems to improve the robustness against adversarial attacks. 
For this purpose, we make the following contributions:
\begin{enumerate}[topsep=3pt, itemsep=1pt, partopsep=4pt, parsep=4pt]
    \item We substitute the ASR system's standard feed-forward NN (fNN) with different network architectures, which are capable of capturing model uncertainty, namely Bayesian NNs (BNN) \cite{neal1995bayesian}, Monte Carlo (MC) dropout~\cite{dropout} and deep ensembles~\cite{deepensemble}.
    \item We calculate different measures to assess the uncertainty when predicting an utterance. Specifically, 
we measure the entropy, variance, averaged Kullback-Leibler divergence, and the mutual information of the NN outputs.
    \item We train a one-class classifier by fitting a normal distribution on the values w.r.t.~these measure for an exemplary set of benign examples. Adversarial examples can then be detected as outliers of the learned distribution. Compared to previous work, this has the advantage that we do not need any adversarial examples to train the classifier and are not tailored to specific kinds of attacks. 
\end{enumerate} 

The results show that we are able to detect adversarial examples with an area under the receiver operating characteristic curve score of more than 0.99 using the NNs output entropy. 
Additionally, the NNs used for uncertainty quantification are less vulnerable to adversarial attacks when compared to a standard feed-forward neural network. The code is available at \url{github.com/rub-ksv/uncertaintyASR}.

\section{Background}
\label{sec: background}
In the following, we briefly outline the estimation of adversarial examples for hybrid ASR systems and introduce a set of approaches for uncertainty quantification in neural networks.

\subsection{Adversarial Examples}
For simplicity, we assume that the ASR system can be written as a function $f$, which takes an audio signal $x$ as input and maps it to its most likely transcription $f(x)$, which should be consistent or at least close to the real transcription $y$. Adversarial examples are a modification of $x$, where specific minimal noise $\delta$ is added to corrupt the prediction, i.\,e., to yield $f(x+ \delta) \neq f(x)$. 

In this general setting, the calculation of adversarial examples for ASR systems can be divided into two steps:

\textbf{Step 1: Forced Alignment.} Forced alignment is typically used for training hybrid ASR systems if no exact alignments between the audio input and the transcription segments are available. The resulting alignment can be used to obtain the NN output targets for Step~2. Here, we utilize the forced alignment algorithm to find the best possible alignment between the original audio input and the malicious target transcription.

\textbf{Step 2: Projected Gradient Descent.}
In this paper, we use the projected gradient descent (PGD) method to create adversarial examples for the targets derived in Step~1. PGD finds solutions 
by gradient descent, i.e., by iteratively computing the gradient of a loss with respect to $\delta$ and moving into this direction. To remain in the allowed perturbation space, $\delta$ is constrained to remain below a pre-defined maximum 
perturbation~$\epsilon$. 

\subsection{Neural Networks for Uncertainty Quantification}

A range of approaches have recently been proposed for quantifying uncertainty in NNs: 

\textbf{Bayesian Neural Networks:} A mathematically grounded method for quantifying uncertainty in neural networks is given by Bayesian NNs (BNNs)~\cite{neal1995bayesian}. Central to these methods is the calculation of a posterior distribution over the network parameters, which models the probabilities of different prediction networks. The final predictive function is derived as 
\begin{equation}\label{BNN}
 p(y|x, \mathcal{D}) = \int p(y|x, \theta)p(\theta|\mathcal{D})d\theta \enspace,
\end{equation}
where  $p(\theta | \mathcal{D})$ is the posterior distribution of the parameters $\theta$, $y$ the output, $x$ the input and $\mathcal{D}=\{(x_i, y_i)\}_{i=1}^n$ the training set. To approximate the often intractable posterior distribution, variational inference methods can be applied. These fit a simpler distribution $q(\theta | D)$ as close to the true posterior as possible by minimizing their Kullback-Leibler divergence (KLD). Minimizing this, again intractable, KLD is equal to maximizing the so-called \textit{evidence lower bound} (ELBO) given by 
\begin{equation}\label{ELBO}
    \mathbb{E}_{q(\theta | D)}[\log p(y_i|x_i ,\theta)] - \text{KLD}[q(\theta|D)||p(\theta)] \enspace .
\end{equation}
During prediction, the integral of Eq.~\eqref{BNN} is approximated by averaging  $p(y|x, \theta_t)$ for multiple samples $\theta_t$ drawn from $q(\theta | D)$.

While there are different approaches to BNNs, we follow Louizos~\etal{} \cite{vmg_BNN} in this paper.

\textbf{Monte Carlo Dropout:} Another approach that scales to deep NN architectures is Monte Carlo dropout \cite{dropout}, which was introduced as an approximation to the Bayesian inference. In this approach, the neurons of an NN are dropped with a fixed probability during training and testing. This can be seen as sampling different sub-networks consisting of only a subset of the neurons and leading to different prediction results for the same input. Here $\theta_t$ denotes the model parameters for the $t$-th sub-network and the final prediction is given by $p(y|x) =  \frac{1}{T} \sum_{t=1}^T  p(y|x, \theta_t)$.

\textbf{Deep Ensembles:} A simple approach, which has been found to often outperform more complex ones \cite{Nowozin}, is the use of a deep ensemble \cite{deepensemble}. The core idea is to train multiple NNs with different parameter initializations on the same data set. In this context, we denote the prediction result of the $t$-th NN by $p(y|x, \theta_t)$. The final prediction is again given by the average over all $T$ model $p(y|x) =  \frac{1}{T} \sum_{t=1}^T  p(y|x, \theta_t)$.

\section{Approach}
\label{sec:approach}

For the detection of the attack, \ie~the identification
of adversarial examples, we describe the general attack setting and the different uncertainty measures that we employ.

\subsection{Threat Model}
We assume a white-box setting in which the attacker has full access to the model, including all parameters. Using this knowledge, the attacker generates adversarial examples offline. We only consider targeted attacks, where the adversary chooses the target transcription. Additionally, we assume that the trained ASR system remains unchanged over time.
 
\subsection{Uncertainty Measures}
For quantifying prediction uncertainty, we employ the following measures:

\label{sec:measures}
\textbf{Entropy:} To measure the uncertainty of the network over class predictions, we calculate the \textit{entropy} over the $K$ output classes as
\begin{equation}
\mathcal{H}[p(y|x)] = -\sum_{c=1}^{K}  p(y_c|x) \cdot \log p(y_c|x) \enspace.
\end{equation}
This can be done for all network types, including the fNN with a softmax output layer. We calculate the entropy for each time step and use its maximum value as the uncertainty measure. 

\textbf{Mutual Information:} 
To leverage the possible benefits of replacing the fNN with a BNN, MC dropout, or a deep ensemble, we evaluate the multiple predictions $p(y|x, \theta_t)$ for $t=1,...,T$ of these networks. Note that these probabilities are derived differently for each network architecture, as described in Section~\ref{sec: background}. With this setup we can calculate the \textit{mutual information} (MI), which is upper bounded by the entropy and defined through 
\begin{equation}\label{mi}
    \text{MI}= \mathcal{H}[p(y|x)]  - \frac{1}{T}\sum_{t=1}^T \mathcal{H}[p(y|x, \theta_t)] \enspace .
\end{equation}
The MI indicates the inherent uncertainty of the model on the presented data~\cite{prior_networks}.

\textbf{Variance:} Another measure that has been used by Feinman~\etal~\cite{feinman2017detecting} to detect adversarial examples for image recognition tasks is the \textit{variance} of the different predictions:
\begin{equation}
\frac{1}{T}\sum_{t=1}^T p(y|x, \theta_t)^2 - p(y|x)^2 \enspace .
\end{equation}

\textbf{Averaged Kullback-Leibler Divergence:} To observe the variations of the distributions---without the mean reduction used for the variance---we further introduce the \textit{averaged Kullback-Leibler divergence} (aKLD). It is defined as
\begin{equation}
\frac{1}{T-1}\sum_{t=1}^{T-1} p(y|x, \theta_t) \cdot \log \frac{p(y|x, \theta_t)}{p(y|x, \theta_{t+1})} \enspace .
\end{equation} 
Because  the samples $\theta_t$ are drawn independently, we compare the first drawn example to the second, the second to the third, and so on without any reordering. 
\section{Experiments}

In the following, we give implementation details and describe the results of our experimental analysis. 

\label{sec:results}
\subsection{Recognizer}

We use a hybrid deep neural network - hidden Markov model ASR system. As a proof of concept for adversarial example detection, we focus on a simple recognizer for sequences of digits from 0 to 9. The code is available at \url{github.com/rub-ksv/uncertaintyASR}.

We train the recognizer with the \emph{TIDIGITS} training set, which includes approximately 8000 utterances of digit sequences. The feature extraction is integrated into the NNs via \emph{torchaudio}. We use the first 13 mel-frequency cepstral coefficients (MFCCs) and their first and second derivatives as input features and train the NNs for 3 epochs followed by 3 additional epochs of Viterbi training to improve the ASR performance.

We use NNs with two hidden layers, each with 100 neurons, and a softmax output layer of size 95, corresponding to the number of states of the hidden Markov model (HMM). For the deep ensemble, we train $T = 5$ networks with different initialization; for the BNN, we draw $T=5$ models from the posterior distribution and average the outputs to form the final prediction; 
and for dropout, we sample $T=100$ sub-networks for the average prediction.\footnotemark \footnotetext{Note, that we needed to increase the number of samples for dropout compared to the other methods, since using $T=5$ for dropout led to worse recognition accuracy. Moreover, we also needed to estimate the average gradient over 10 sub-nets per training sample during training to observe increased robustness against adversarial examples.}

The ASR accuracies are evaluated on a test set of 1000 benign utterances and are shown in Table~\ref{tab:benign}, calculated as the sum over all substituted words $S$, inserted words $I$, and deleted words $D$ in comparison to the original and the target label
\begin{equation}
    \text{Accuracy}= \frac{N -I -D - S}{N} \enspace ,
\end{equation}
where $N$ is the total number of words of the reference text, either the original or the malicious target text.

All methods lead to a reasonable accuracy, with the deep ensemble models outperforming the fNN. At the same time, there is some loss of performance for the MC dropout model and the BNN model.

\begin{table}
\caption{Accuracy on benign examples.}
\smallskip
\centering
\begin{tabular}{cccc} 
\toprule
fNN & deep ensemble & MC dropout & BNN\\ 
\midrule 
0.991 & 0.994 & 0.973 & 0.981 \\
\bottomrule
\end{tabular}
\label{tab:benign}
\end{table}

\subsection{Adversarial Attack}
For the attack, we use a sequence of randomly chosen digits with a random length between 1 and 5. The corresponding targets for the attack have been calculated with the Montreal forced aligner~\cite{mcauliffe-2017-montreal}. To pass the targets through the NN we used the projected gradient descent (PGD) attack \cite{Madry}. For this purpose, we used cleverhans, a Python library to assess machine learning systems against adversarial examples~\cite{papernot-2018-cleverhans}. 

During preliminary experiments, we found that using multiple samples for estimating the stochastic gradient for the estimation of adversarial examples decreases the strength of the attack. This result contradicts insights found for BNNs in image classification tasks, where the adversarial attacks become stronger when multiple samples are drawn for the gradient \cite{Handtuch}. An explanation for this finding could be that for image classification, no hybrid system is used. In contrast to that, the Viterbi decoder in a hybrid ASR exerts an additional influence on the recognizer output and favors cross-temporal consistency.

\begin{figure}[h]
\centering
\includegraphics[width=\columnwidth]{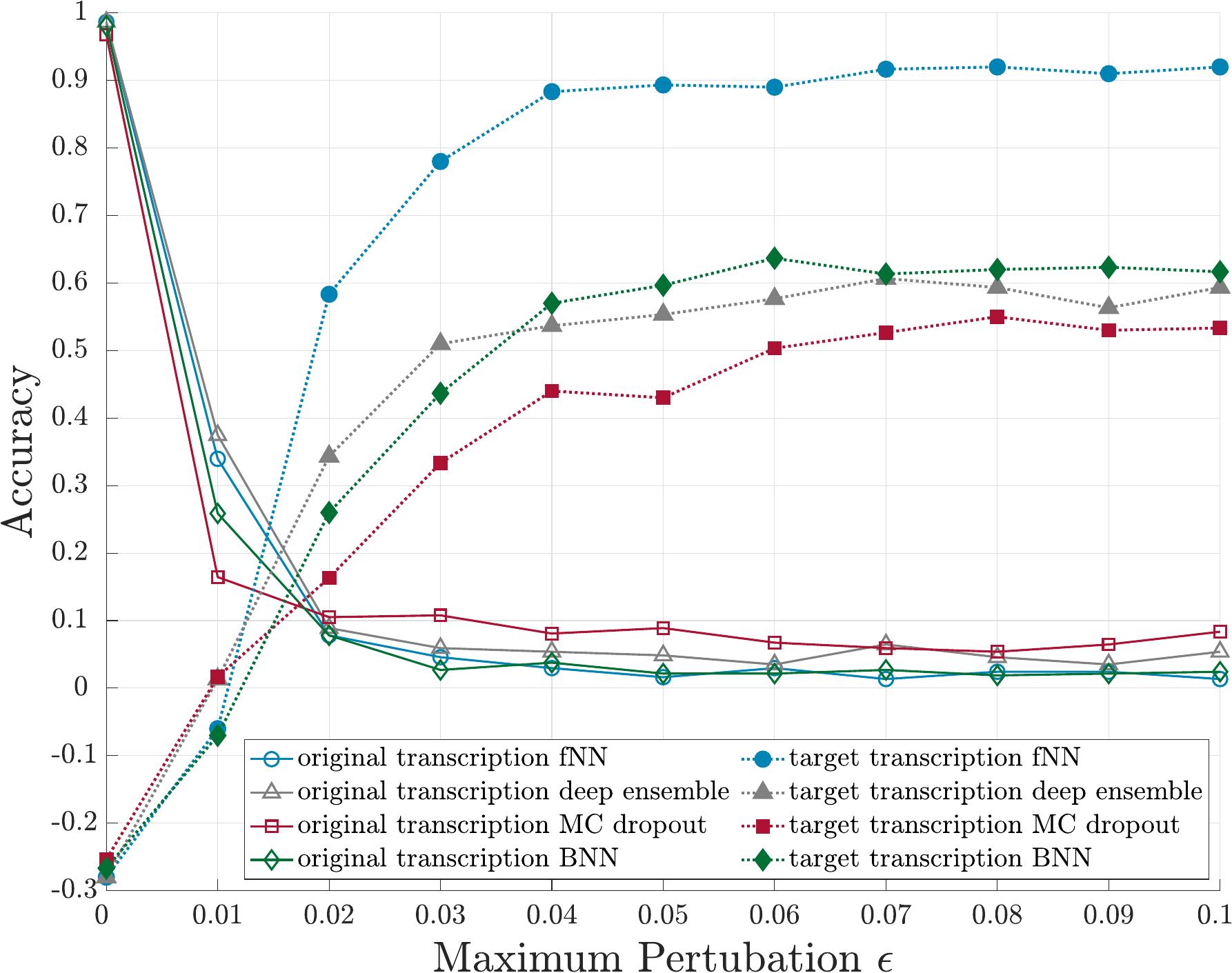}
\caption{Accuracy with respect to the original and the target transcription plotted over $\epsilon$ for fNN, MC dropout, BNN, and deep ensemble, evaluated on 100 utterances each.}
\label{fig:accuracy}
\end{figure}

Correspondingly, our empirical results indicate that sampling multiple times leads to unfavorable results for ASR from the attacker's perspective. Evaluating the averaged and the single adversarial examples separately shows that the averaged adversarial examples are more likely to return the original text due to the Viterbi decoding of the hybrid ASR system. Consequently, we have only used one sample to improve the attacker's performance and, thus, evaluate our defense mechanisms against a harder opponent.

To validate the effectiveness of PGD, we investigate the word accuracy of the label predicted for the resulting adversarial example w.r.t.~the target and the original transcription. These word accuracies are shown in Figure~\ref{fig:accuracy} for varying perturbation strength ($\epsilon = 0, \dots, 0.1 $ with a step size of 0.01) of PGD attack. Note that $\epsilon = 0$ corresponds to benign examples, as no perturbations are added to the original audio signal. We evaluated 100 adversarial examples for each $\epsilon$ and NN.
 
For all models, the accuracy w.r.t.~the target transcription increases with increasing perturbation strength until approximately $\epsilon=0.04$, and stagnates afterward. The attack has the most substantial impact on the fNN-based model, where the accuracy w.r.t.~the malicious target transcription for $\epsilon\geq0.05$ is almost 50\,\% higher than for the other models, where the accuracy only reaches values between $0.4$ and $0.7$. This indicates that including NNs for uncertainty quantification into ASR systems makes it more challenging to calculate effective targeted adversarial attacks.
Nevertheless, the accuracy w.r.t~the original transcription is equally affected across all systems, indicating that for all of them, the original text is difficult to recover under~attack.

\subsection{Classifying Adversarial Examples}
\begin{figure}[t]
\centering
\includegraphics[width=0.95\columnwidth]{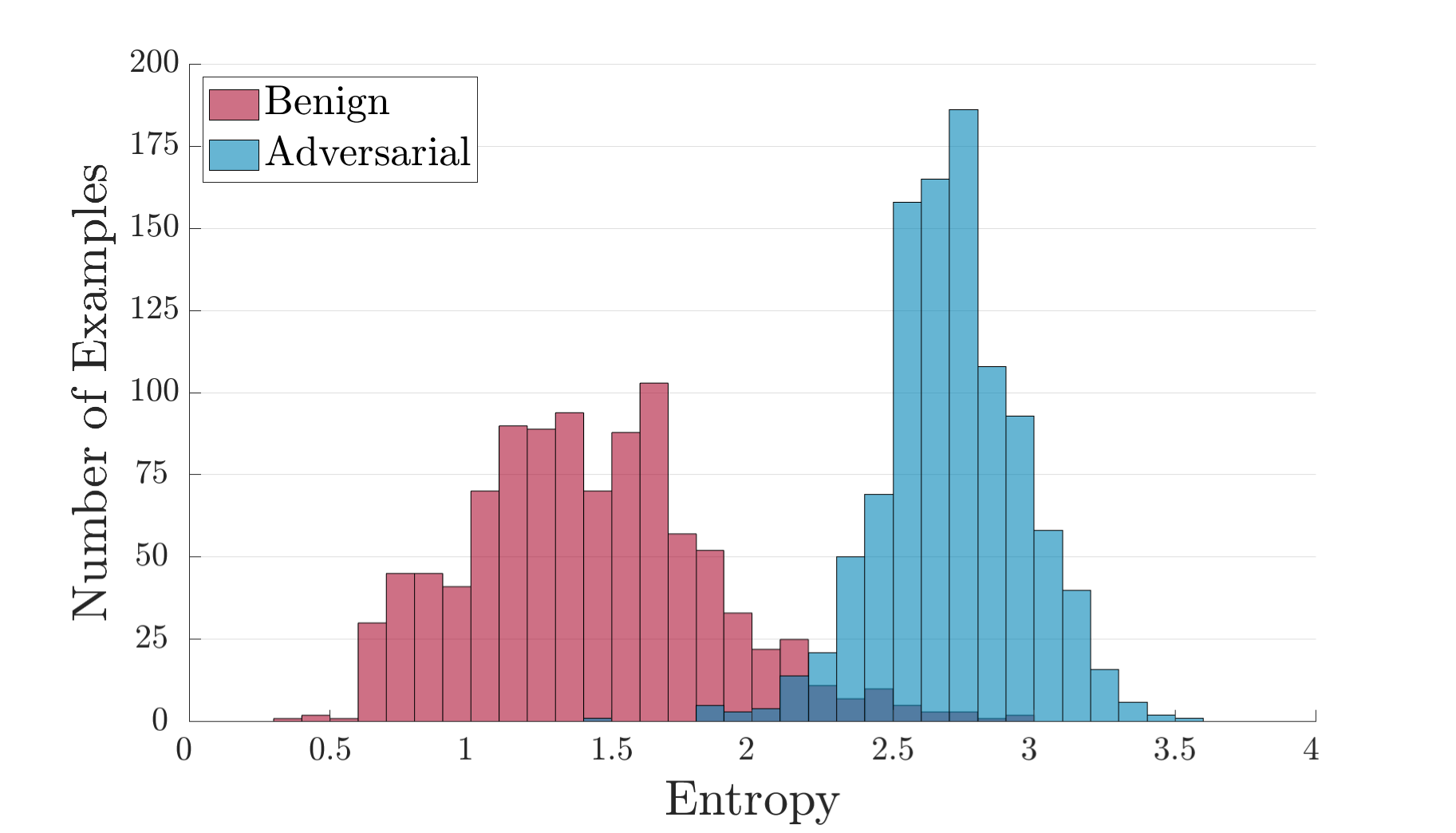}
\caption{Histograms of predictive entropy values for an fNN for 1000 benign and 1000 adversarial examples.} 
\label{fig:Histogram}
\end{figure}

In order to detect adversarial examples, we calculate the measures described in Section~\ref{sec:measures} for 1000 benign and 1000 adversarial examples, estimated via PGD with $\epsilon = 0.05 $.
Figure~\ref{fig:Histogram} exemplary shows histograms of the entropy values of the predictive distribution of the fNN over both sets of examples. Like the fNN, all other models also clearly tend to display higher uncertainty over classes for adversarial examples, while the difference between benign and adversarial examples was most severe for the entropy.

\begin{figure}
\centering
\includegraphics[width=0.95\columnwidth]{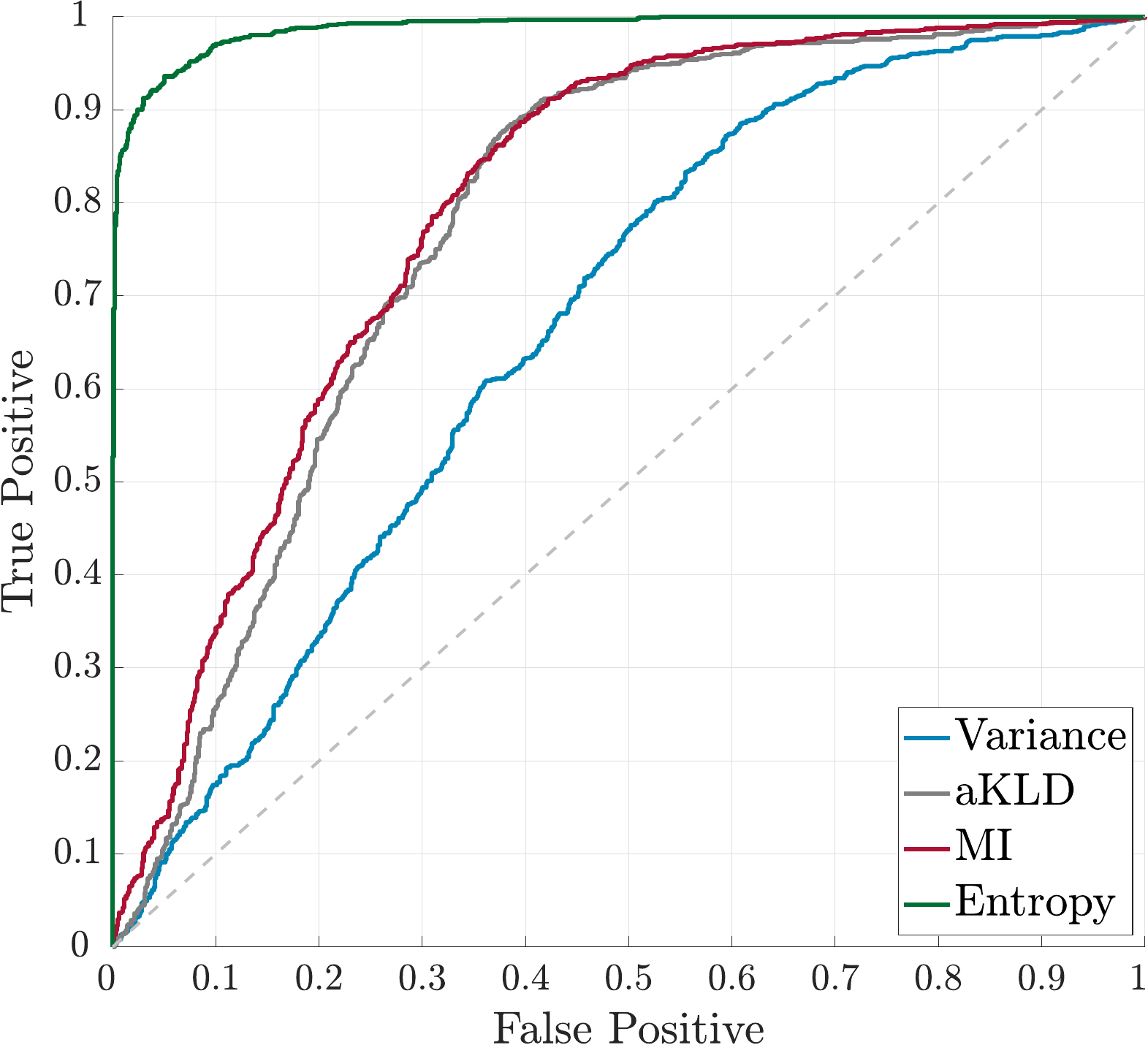}
\caption{ROC curves of the different measures for the BNN with $\epsilon = 0.05$ on 1000 benign and adversarial examples each.}
\label{fig:ROC}
\end{figure}

We build on this observation by constructing simple classifiers for the detection of adversarial examples: 
We fit a Gaussian distribution to the values of the corresponding measure over a held-out data set of 1000 benign examples for each network and measure. A new observation can then be classified as an attack if the value of the prediction uncertainty has low probability under the Gaussian model.
We measure the receiver operating characteristic (ROC) of these classifiers for each model type and uncertainty measure. The results are shown exemplarily for the BNN in Figure~\ref{fig:ROC}. Additionally, we display the area under the ROC curve (AUROC) in Table~\ref{tab:AUROC0.05}. The results show that only the entropy has stable performance across all kinds of NNs and clearly outperforms the other measures (variance, aKLD, and MI). Note that the entropy is also the only measure that can be calculated for the fNN. 

\begin{table}[h!]
\caption{AUROC feature scores for 1000 adversarial examples with a perturbation strength $\epsilon = 0.05$. Best results for each network are shown in bold.}
\smallskip
\centering
\begin{tabular}{l|ccc|c} 
\toprule
  &  Variance & aKLD & MI &Entropy \\ 
\midrule 
fNN & -- & -- & -- & \textbf{0.989}   \\
deep ensemble & 0.455 & 0.892 & \textbf{0.993} & 0.990  \\
MC dropout & 0.637 & 0.443 & 0.498 & \textbf{0.978} \\
BNN & 0.667 & 0.777 & 0.794 & \textbf{0.988}  \\
\bottomrule
\end{tabular}
\label{tab:AUROC0.05}
\end{table}

\begin{table}[h!]
\caption{AUROC feature scores for 1000 adversarial examples with a perturbation strength $\epsilon = 0.02$. Best results for each network are shown in bold.}
\smallskip
\centering
\begin{tabular}{l|ccc|c} 
\toprule
  &  Variance & aKLD & MI &Entropy \\ 
\midrule 
fNN & -- & -- & -- & \textbf{0.997}   \\
deep ensemble & 0.461 & 0.624 & 0.964 & \textbf{0.996}  \\
MC dropout & 0.937 & 0.578 & 0.411 & \textbf{0.991} \\
BNN & 0.489 & 0.448 & 0.462 & \textbf{0.998}  \\
\bottomrule
\end{tabular}
\label{tab:AUROC0.02}
\end{table}

To verify the results for adversarial examples with low perturbations, which might be harder to detect, we followed the same approach for 1000 adversarial examples with a maximal perturbation of $\epsilon = 0.02$. The results, shown in Table~\ref{tab:AUROC0.02}, are similar to the ones with the higher perturbation.

\section{Discussion \& Conclusions}
\label{sec:discussion}

Our empirical results show that in a hybrid speech recognition system, replacing the standard feed-forward neural network by a Bayesian neural network, Monte Carlo dropout, or deep ensemble networks increases the robustness against targeted adversarial examples tremendously. This can be seen in the low accuracy of the target transcription, which indicates a far lower vulnerability than that of standard hybrid speech recognition.

Another finding of this work is that the entropy serves as a good measure for identifying adversarial examples. In our experiments, we were able to discriminate between benign and adversarial examples with an AUROC score of up to 0.99 for all network architectures. Interestingly, the other measures which are available when using approaches especially designed
for uncertainty quantification did not improve upon these results.

In future research, it would be interesting to evaluate this setting on a large-vocabulary speech recognition system, to see if (an expected) qualitative difference appears between the networks.

\bibliographystyle{IEEEtran}

\bibliography{mybib}
\end{document}